\documentclass{revtex4}

\usepackage[usenames,dvipsnames]{color}
\usepackage{amsfonts,amssymb,amsmath,amsthm}
\usepackage{bm}
\usepackage[a4paper]{geometry}
\usepackage{pdflscape}
\usepackage{graphicx}
\usepackage{mathrsfs}

\definecolor{darkred}{rgb}{.8,0,0}

\definecolor{darkblue}{rgb}{0,0,.7}

\newcommand{\eps}{\varepsilon}

\newcommand{\rev}[1]{\widetilde{#1}}
\newcommand{\blade}[2]{{#1}_1 \wedge \ldots \wedge {#1}_{#2}}
\newcommand{\bladecheck}[3]{{#1}_1 \wedge \ldots \wedge \check{#1}_{#3} \wedge \ldots \wedge {#1}_{#2}}

\newcommand{\veclist}[2]{{#1}_1,\ldots,{#1}_{#2}}
\newcommand{\uline}[1]{\underline{#1}}
\newcommand{\oline}[1]{\overline{#1}}

\theoremstyle{plain}
\newtheorem*{TrGaugedHam}{Transformation of gauged Hamiltonian}
\newtheorem*{CanEOMGaugedHam}{Canonical equations for gauged Hamiltonian}

\begin{document}

\title{Classical field theories from Hamiltonian constraint: \\
Local symmetries and static gauge fields}

\author{V\'{a}clav Zatloukal}

\email{zatlovac@gmail.com}

\homepage{http://www.zatlovac.eu}

\affiliation{\vspace{3mm}
Faculty of Nuclear Sciences and Physical Engineering, Czech Technical University in Prague, \\
B\v{r}ehov\'{a} 7, 115 19 Praha 1, Czech Republic \\
}

\affiliation{
Max Planck Institute for the History of Science, Boltzmannstrasse 22, 14195 Berlin, Germany
}

\begin{abstract}
We consider the Hamiltonian constraint formulation of classical field theories, which treats spacetime and the space of fields symmetrically, and utilizes the concept of momentum multivector. The gauge field is introduced to compensate for non-invariance of the Hamiltonian under local transformations. It is a position-dependent linear mapping, which couples to the Hamiltonian by acting on the momentum multivector. We investigate symmetries of the ensuing gauged Hamiltonian, and propose a generic form of the gauge field strength. In examples we show how a generic gauge field can be specialized in order to realize gravitational and/or Yang-Mills interaction. Gauge field dynamics is not discussed in this article.

Throughout, we employ the mathematical language of geometric algebra and calculus.
\end{abstract}

\maketitle

%
%

\section{Introduction}

The \emph{Hamiltonian constraint} is a concept useful for Hamiltonian formulation not only of general relativity \cite{DeWitt}, but, in fact, of a generic field theory, as pointed out in Ref.~\cite[Ch.~3]{RovelliQG}, and exploited further in \cite{ZatlCanEOM} and \cite{ZatlSymConsLaws}. Characteristic features of this formulation are: finite-dimensional configuration space, and multivector-valued momentum variable. In this respect it is congruent with the \emph{covariant} (or \emph{De Donder-Weyl}) Hamiltonian formalism \cite{DeDonder,Weyl,GotayI, Echeverria,Leon, Helein2011,Sardanashvily,Kanat1998,Struckmeier} and should be contrasted with the \emph{canonical} (or \emph{instantaneous}) Hamiltonian formalism \cite{GotayII}, which utilizes an infinite-dimensional space of field configurations defined at a given instant of time. Nonetheless, unlike the traditional covariant approaches, the Hamiltonian constraint formalism does not a priori distinguish between spacetime and field variables, leading to a simple, but, at the same time, rather general theory expressed in terms of symmetric and compact equations. Eventually, if the need arises, the full set of variables can be split into the spacetime and field-space component, and the equations can be expressed in terms of functions that depend on spacetime location, which is the usual point of view in field theory. 

Following Chapter~3 in Ref.~\cite{RovelliQG}, let us introduce some basic concepts and terminology of the Hamiltonian constraint formulation of classical field theories, and summarize the results of \cite{ZatlCanEOM} and \cite{ZatlSymConsLaws} relevant for the present article. (In fact, analogous results had been obtained previously in \cite{Helein2002} with an approach based on the \emph{pataplectic} differential form.) 

The \emph{configuration space} $\mathcal{C}$ is a $D+N$-dimensional space of points $q$, which represent possible joint outcomes of \emph{partial observables}, i.e., all physical variables of the theory. (In a typical example of scalar field theory, the partial observables are $D$ spacetime coordinates $x^\mu$ and $N$ real field components $\phi_a$.) \emph{Motions} are $D$-dimensional surfaces $\gamma$ embedded in $\mathcal{C}$, which generalize one-dimensional trajectories of classical particle mechanics. Often, they can be regarded as graphs of functions $\phi_a(x^\mu)$. We shall assume, for simplicity, that $\mathcal{C}$ is a flat (pseudo)Euclidean space. (Note that although in articles \cite{ZatlCanEOM,ZatlSymConsLaws} $\mathcal{C}$ was assumed to be Euclidean, the general results hold in pseudoEuclidean spaces too.)

The \emph{physical} (or \emph{classical}) motions are denoted by $\gamma_{\rm cl}$. For a fixed boundary $\partial\gamma$, they extremize the action functional
\begin{equation} \label{ActionAugm}
\mathcal{A}[\gamma,P,\lambda] 
= \int_\gamma \left[ P(q) \cdot d\Gamma(q) - \lambda(q) H(q,P(q)) \right] ,
\end{equation}
leading to the \emph{canonical equations of motion}
\begin{subequations} \label{CanEOM}
\begin{align} 
\label{CanEOM1}
\lambda \, \partial_P H(q,P) &= d\Gamma , 
\\ \label{CanEOM2}
(-1)^D \lambda \, \dot{\partial}_q H(\dot{q},P) 
&= \begin{cases}
d\Gamma \cdot \partial_q P & ~~{\rm for}~ D=1 \\
(d\Gamma \cdot \partial_q) \cdot P &  ~~{\rm for}~ D>1 ,
\end{cases}
\\
\label{CanEOM3} 
H(q,P) &= 0 .
\end{align}
\end{subequations}
Here, $d\Gamma$ is the infinitesimal oriented element of the surface $\gamma$, $P$ is the $D$-vector-valued momentum field, and $\lambda$ is a (scalar) Lagrange multiplier corresponding to the \emph{Hamiltonian constraint}~(\ref{CanEOM3}). $H$ is a generic function of $q$ and $P$, the so-called relativistic Hamiltonian (or simply Hamiltonian), which characterizes the dynamics of the physical system in question.

Let $q'=f(q)$ be a diffeomorphism on the configuration space $\mathcal{C}$, which transforms  the relevant quantities as follows:
\begin{equation} \label{TrRules}
d\Gamma'=\uline{f}(d\Gamma) \quad,\quad
P'=\oline{f^{-1}}(P) \quad,\quad
\lambda'(q')=\lambda(q) \quad,\quad
H'(q',P') = H(q,P) .
\end{equation}
(The definition of induced mappings $\uline{f}$ and $\oline{f^{-1}}$ is recalled in Appendix~\ref{sec:GAGCTrInd}).
A physical motion $\gamma_{\rm cl}$ is mapped by $f$ to 
\begin{equation} \label{TrPhysMotion}
\gamma_{\rm cl}'=\{f(q)\,|\,q \in \gamma_{\rm cl} \} \equiv f(\gamma_{\rm cl}) ,
\end{equation}
which is a physical motion of the transformed Hamiltonian $H'$.
This is a consequence of the action (\ref{ActionAugm}) being equal for primed and unprimed quantities.

If $H$ and $H'$ coincide, i.e, if
\begin{equation} \label{SymCrit}
H(f(q),\overline{f^{-1}}(P;q)) = H(q,P) \quad(\forall q,P) ,
\end{equation}
then physical motions are mapped to physical motions of the same system, and $f$ is said to be a \emph{symmetry}. For infinitesimal transformations $f(q) = q + \eps\, v(q)$, $\eps \ll 1$, determined by a vector field $v$, Eq.~(\ref{SymCrit}) takes the form
\begin{equation} \label{SymCritInfsm}
v \cdot \dot{\partial}_q H(\dot{q},P)
- \big( \dot{\partial}_q \wedge (\dot{v} \cdot P) \big) \cdot \partial_P H(q,P)
= 0 .
\end{equation}

Now, the canonical equations (\ref{CanEOM1}) and (\ref{CanEOM2}) can be combined to yield
\begin{equation} \label{PreConsLaw}
(-1)^{D} \lambda \left[ v \cdot \dot{\partial}_q H(\dot{q},P)
- \big( \dot{\partial}_q \wedge (\dot{v} \cdot P) \big) \cdot \partial_P H(q,P) \right]
= \begin{cases}
d\Gamma \cdot \partial_q (P \cdot v) & ~~{\rm for}~ D=1 \\
(d\Gamma \cdot \partial_q) \cdot (P \cdot v) &  ~~{\rm for}~ D>1 ,
\end{cases}
\end{equation}
and so we observe that if the infinitesimal symmetry condition~(\ref{SymCritInfsm}) is fulfilled, the left-hand side vanishes, and Eq.~(\ref{PreConsLaw}) expresses a conservation law for the quantity $P \cdot v$. So much for the results of our previous articles \cite{ZatlCanEOM} and \cite{ZatlSymConsLaws}.

In this article, we will be concerned with the situation when the Hamiltonian of the system is invariant under some class of global (rigid) transformations, but fails to be invariant under more generic \emph{local} transformations that possess additional position dependence. To compensate for this non-invariance, we introduce in Sec.~\ref{sec:StatGaugeField} a \emph{gauge field} with appropriate transformation properties, which acts as a $q$-dependent linear map on the momentum multivector $P$. The ensuing gauged Hamiltonian acquires thereupon an additional $q$-dependency, and the canonical equations~(\ref{CanEOM}) are modified to Eqs.~(\ref{CanEOMgauge}). 

The structure of the latter canonical equations suggests to define, by Eq.~(\ref{FieldStrength}) in Sec.~\ref{sec:GaugeFieldStr}, the \emph{field strength} corresponding to the gauge field, which is a linear $q$-dependent function that maps grade-$r$ multivectors to grade-$r+1$ multivectors. In Appendix~\ref{sec:FieldStrTorsion}, it is interpreted as torsion corresponding to the Weitzenb\"{o}ck connection on the configuration space $\mathcal{C}$ . 

In the present article, the gauge field and the field strength are static background quantities in the sense that they do not obey their own dynamical equations of motion, but rather are prescribed by some external body. This is also why we do not attempt to include them in the set of partial observables, but treat them separately. We relegate the study of gauge field dynamics (presumably implemented by means of a suitable kinetic term) to the future.

Conservation laws maintain the form of Eq.~(\ref{PreConsLaw}), where the left-hand side features the gauged Hamiltonian. Whether a vector field $v$ is a symmetry generator thus depends on the concrete form of the gauge field. The issue is discussed in Sec.~\ref{sec:SymGaugedHam}.

The general theory of Sections~\ref{sec:StatGaugeField}, \ref{sec:GaugeFieldStr} and \ref{sec:SymGaugedHam} is much illuminated through the examples of Sec.~\ref{sec:Examples}. In Examples~\ref{sec:ExGr} and \ref{sec:ExYM}, without any reference to a concrete form of the Hamiltonian, the partial observables are divided into $D$ spacetime coordinates and $N$ field components, and two subgroups of the group of all configuration-space diffeomorphisms are considered. In the first example, we ``gauge" spacetime diffeomorphisms by a gauge field equivalent to the tetrad (or vierbein) in the tetrad formulation of gravity. It acts nontrivially only on the spacetime, and therefore has fewer degrees of freedom than a generic gauge field. In the second example, we consider rotations in the field space that depend on spacetime location. This leads to the Yang-Mills gauge field characterized by a bivector-valued gauge potential $A_\mu$. 

In Example~\ref{sec:ExSF}, we choose the Hamiltonian of a scalar field theory coupled to a generic gauge field, which is subsequently specialized, respectively, to the gravitational field of Example~\ref{sec:ExGr}, and the Yang-Mills field (in particular, electromagnetic field) of Example~\ref{sec:ExYM}. In this sense, the generic gauge field unifies gravitational and Yang-Mills fields. 

In the last example, Sec.~\ref{sec:ExStrGr}, the configuration space $\mathcal{C}$ is identified with a $D+N$-dimensional spacetime, in which point particles, strings, or higher-dimensional membranes (depending on the value of $D$) propagate. Complete group of spacetime diffeomorphisms is gauged by a gauge field that thus embodies the gravitational field. Relation to standard metric formulation of gravity is discussed in detail. Namely, we recover the traditional form of the geodesic equation, and notice that the symmetry condition, Eq.~(\ref{SymInfsmGauge}), is equivalent with the celebrated Killing equation, where the symmetry generator $v$ is the Killing vector. 

The mathematical formalism we use is somewhat uncommon, but proves to be very convenient when it comes to handling higher-dimensional geometric objects such as the motions $\gamma$ with their surface elements $d\Gamma$, the momentum multivector $P$, etc. It is more explicit and versatile than the language of differential forms, and, at the same time, maintains coordinate freedom, so that expressions can be written in a succinct form without the need to introduce multiple indices (what is the case in tensor calculus). It goes under the name \emph{geometric} (or \emph{Clifford}) \emph{algebra and calculus}, and we follow its exposition as provided in Refs.~\cite{Hestenes} and \cite{DoranLas}. Although we recall some important definitions and results in Appendix~\ref{sec:GAGC}, we do not attempt to supply a complete self-contained presentation of geometric algebra and calculus in this article. In order to fully understand all manipulations that follow, the reader is encouraged to consult the above monographs, or, for concise introduction, appendices of our previous articles \cite{ZatlCanEOM,ZatlSymConsLaws}.

\section{Static gauge field}
\label{sec:StatGaugeField}

%

We have seen that a transformation $f:q\mapsto q'$ of the configuration space $\mathcal{C}$ is a symmetry of Hamiltonian $H$ if Eq.~(\ref{SymCrit}) is satisfied. It often happens that
\begin{equation} \label{GlobSym}
H(q',P) = H(q,P) .
\end{equation}
(For example, the physical system does not depend on certain partial observable or a combination of those.) But this equation by itself does not imply Eq.~(\ref{SymCrit}) as one has to take into account also the transformation of momentum, $P'=\oline{f^{-1}}(P;q)$, which depends on derivatives of $f$, and so can become rather complicated for \emph{local} transformations $f$, i.e., those that vary from point to point. Therefore, roughly speaking, the more generic transformations we consider, the less likely it is that they will be symmetries of the Hamiltonian.

Nevertheless, we can impose that a certain transformation, or a class of transformations, $f$ be a symmetry of our system as follows.
Consider a modified \emph{gauged Hamiltonian}
\begin{equation} \label{HamGauged}
H_h(q,P) = H(q,\oline{h}(P;q)) ,
\end{equation}
where $\oline{h}$ is a $q$-dependent linear outermorphism (see Appendix~\ref{sec:GAGCLinFunc}), the so-called \emph{gauge field}, that maps momentum multivectors $P$ to $\oline{h}(P)$. If $\gamma_{\rm cl}$ is a physical motion with respect to $H_h$, then the motion $\gamma'_{\rm cl}$ defined by Eq.~(\ref{TrPhysMotion}) is a physical motion with respect to a new Hamiltonian $H_h'$, related to $H_h$ by Eq.~(\ref{TrRules}). Assuming that Eq.~(\ref{GlobSym}) holds, we find
\begin{equation}
H_h'(q',P') 
= H_h(q,P) 
= H(q,\oline{h}(P))  
= H(q',\oline{h}\,\oline{f}(P')) 
= H(q',\oline{h'}(P'))
= H_{h'}(q',P') ,
\end{equation}
where we have defined the transformation rule for the gauge field
\begin{equation} \label{TrGaugeField}
\oline{h'}(b;q') = \oline{h}\big(\oline{f}(b;q);q \big)
\quad,\quad \forall ~{\rm vectors}~ b .
\end{equation}
To summarize, we obtain the following proposition.
\begin{TrGaugedHam}
Let $f:\mathcal{C}\rightarrow\mathcal{C}$ be an arbitrary diffeomorphism, and suppose that the Hamiltonian $H$ is such that Eq.~(\ref{GlobSym}) holds. If $\gamma_{\rm cl}$ is a physical motion of the gauged Hamiltonian $H_h$, then $f(\gamma_{\rm cl})$ is a physical motion of $H_{h'}$, where the gauge field transforms as $\oline{h'} = \oline{h}\,\oline{f}$.
\end{TrGaugedHam}
This means that $f$ may well be called a symmetry as long as we admit that gauged Hamiltonians whose gauge fields are related by Eq.~(\ref{TrGaugeField}) describe the same physical system. The transformation $f$ is then referred to as the \emph{gauge transformation}.

The gauge field associates to each point of $\mathcal{C}$ a linear map $\oline{h}$, which uniquely determines its adjoint $h$, and the respective inverses $\oline{h^{-1}}$ and $h^{-1}$ (we shall assume that $h$ is invertible). One may encounter the gauge field in any of the four equivalent forms. Their transformation rules can be derived easily from Eq.~(\ref{TrGaugeField}):
\begin{equation} \label{TrGaugeFieldAll}
\oline{h'} = \oline{h} \, \oline{f} \quad,\quad
h'^{-1} = h^{-1} \uline{f}^{-1} \quad,\quad
\oline{h'^{-1}} = \oline{f^{-1}} \, \oline{h^{-1}} \quad,\quad
h' = \uline{f} \, h .
\end{equation}
Being a linear function, the gauge field has, in general, $(D+N)^2$ degrees of freedom. This number can be reduced if we consider only a subgroup of the group of all diffeomorphisms of $\mathcal{C}$, in which case it is sufficient (but not ``obligatory") to assume that $h$ has a certain more restricted form (see Examples~\ref{sec:ExGr} and \ref{sec:ExYM}, where $h$ is reduced, respectively, to the gravitational and Yang-Mills field).

In view of Eq.~(\ref{HamGauged}), the gauged Hamiltonian can be regarded as an ordinary Hamiltonian with extra position dependence due to the gauge field, so the canonical equations of motion, Eqs.~(\ref{CanEOM}), apply to $H_h$ without change. Nevertheless, it is beneficial to express them in terms of the original ungauged Hamiltonian $H$. For this purpose we calculate
\begin{equation} \label{GaugedHamDifP}
\partial_P H_h(q,P) = \partial_P H(q,\oline{h}(P))
= \partial_P \oline{h}(P) \cdot \partial_{\bar{P}} H(q,\bar{P})
= h(\partial_{\bar{P}}) H(q,\bar{P}) ,
\end{equation}
where we have denoted $\bar{P} \equiv \oline{h}(P)$, and
\begin{equation} \label{GaugedHamDifq}
\dot{\partial}_q H_h(\dot{q},P)
= \dot{\partial}_q H(\dot{q},\dot{\oline{h}}(P))
= \dot{\partial}_q H(\dot{q},\bar{P})
+ \dot{\partial}_q \dot{\oline{h}}(P) \cdot \partial_{\bar{P}} H(q,\bar{P}) .
\end{equation}
With these relations in hand, we readily obtain

\begin{CanEOMGaugedHam}
The canonical equations of motion for a gauged Hamiltonian $H_h$, related to the original Hamiltonian $H$ by Eq.~(\ref{HamGauged}), read
\begin{subequations} \label{CanEOMgauge}
\begin{align} 
\label{CanEOM1gauge}
\lambda \, \partial_{\bar{P}} H(q,\bar{P}) &= h^{-1}(d\Gamma) , 
\\ \label{CanEOM2gauge}
(-1)^D \lambda\, \dot{\partial}_q H(\dot{q},\bar{P}) 
+ (-1)^D \dot{\partial}_q \dot{\oline{h}}(P) \cdot h^{-1}(d\Gamma)
&= \begin{cases}
d\Gamma \cdot \partial_q P & ~~{\rm for}~ D=1 \\
(d\Gamma \cdot \partial_q) \cdot P &  ~~{\rm for}~ D>1 ,
\end{cases}
\\
\label{CanEOM3gauge} 
H(q,\bar{P}) &= 0 ,
\end{align}
Alternatively, the second canonical equation (\ref{CanEOM2gauge}) can be cast in terms of $\bar{P} \equiv \oline{h}(P)$ as
\begin{equation} \label{CanEOM2gauge*}
(-1)^D \lambda\, \oline{h}(\dot{\partial}_q) H(\dot{q},\bar{P}) 
- h^{-1}(d\Gamma) \cdot \oline{h} \big( \dot{\partial}_q \wedge \dot{\oline{h^{-1}}} (\bar{P}) \big)
= \begin{cases}
d\Gamma \cdot \partial_q \bar{P} & ~~{\rm for}~ D=1 \\
h^{-1} (d\Gamma \cdot \partial_q) \cdot \bar{P} &  ~~{\rm for}~ D>1 .
\end{cases}
\end{equation}
\end{subequations}
\end{CanEOMGaugedHam}
To derive Eq.~(\ref{CanEOM2gauge*}), we employed identities (\ref{GAadjointIdent}) and (\ref{GAhhDer}) to find
\begin{equation}
(d\Gamma \cdot \partial_q) \cdot P
= (d\Gamma \cdot \partial_q) \cdot \oline{h^{-1}}(\bar{P})
= (d\Gamma \cdot \dot{\partial}_q) \cdot \dot{\oline{h^{-1}}}(\bar{P})
+ \oline{h^{-1}} \big( h^{-1}(d\Gamma \cdot \partial_q) \cdot (\bar{P}) \big) 
\end{equation}
and
\begin{equation}
\dot{\partial}_q \dot{\oline{h}}(P) \cdot h^{-1}(d\Gamma)
= \dot{\partial}_q \oline{h^{-1}} \, \dot{\oline{h}}(P) \cdot d\Gamma
= - \dot{\partial}_q \dot{\oline{h^{-1}}}(\bar{P}) \cdot d\Gamma ,
\end{equation}
and finally rearranged the terms using Eq.~(\ref{GAident1}).

Canonical equations in the form with Eq.~(\ref{CanEOM2gauge*}) exhibit more clearly their invariance with respect to gauge transformation $f$. That is, the functions $(d\Gamma,P,\lambda,h)$ and their counterparts $(d\Gamma',P',\lambda',h')$, transformed according to Eqs.~(\ref{TrRules}) and (\ref{TrGaugeFieldAll}), follow the same differential equations with the same Hamiltonian function $H$ (which is assumed to have the property expressed by Eq.~(\ref{GlobSym})). 
Indeed, observe that individual constituents are \emph{gauge-invariant}, i.e., they transform trivially:
\begin{equation} \label{GaugeInvQuant}
\oline{h'}(P') = \oline{h}(P) \quad,\quad
h'^{-1}(d\Gamma') = h^{-1}(d\Gamma) \quad,\quad 
\oline{h'}(\partial_{q'}) = \oline{h}(\partial_q) ,
\end{equation}
where the vector derivative $\partial_q$ transforms according to Eq.~(\ref{GCTrDeriv}).
Only the second term on the left-hand side of Eq.~(\ref{CanEOM2gauge*}) needs closer inspection (see Sec.~\ref{sec:GaugeFieldStr} below). 

The differential operator $\oline{h}(\partial_q)$ is invariant under $f$, and hence it deserves the name \emph{gauge-invariant derivative}. Its significance has been emphasized in the context of Gauge Theory Gravity \cite{DoranLas,Lasenby1998,Hestenes2005}, where the so-called \emph{displacement gauge field} $\oline{h}$ ensures invariance of the theory under spacetime diffeomorphisms. Here we do not restrict our attention only to spacetime transformations, but allow also for transformations in the field space (or even for those that mix the two spaces), keeping the same generic form of the gauge field $\oline{h}$. It is then instructive to observe, in Example~\ref{sec:ExSF} below, how $\oline{h}$ corresponding to local rotations in the field space gives rise to a coupling of a scalar field theory to the Yang-Mills background field, expressed in terms of the traditional Yang-Mills covariant derivative, Eq.~(\ref{YMCovDer}).

\section{Gauge field strength}
\label{sec:GaugeFieldStr}

Let us draw our attention to the second canonical equation~(\ref{CanEOM2gauge*}), specifically, to the second term on its left-hand side, and define the \emph{field strength} corresponding to gauge field $h$,
\begin{equation} \label{FieldStrength}
F(\bar{P}) \equiv -\oline{h} \big( \dot{\partial}_q \wedge \dot{\oline{h^{-1}}} (\bar{P}) \big)
= \oline{h}(\dot{\partial}_q) \wedge \dot{\oline{h}}(P) .
\end{equation}
(C.f. the definition of displacement-gauge field strength in \cite[Ch.~13.5.2]{DoranLas}.)

\begin{figure} 
\includegraphics[scale=1]{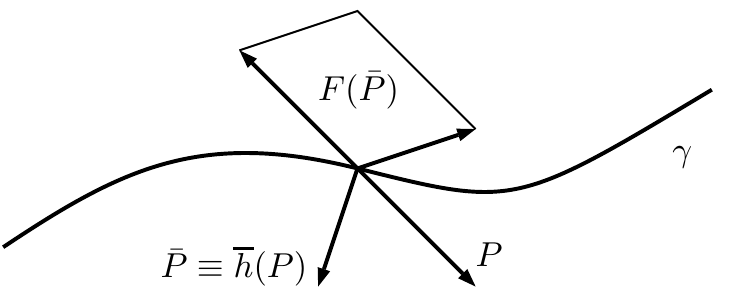}
\caption{Momentum $P$, gauge-invariant momentum $\bar{P}$, and field strength $F(\bar{P})$ for one-dimensional motions $\gamma$.}
\label{fig:FieldStr}
\end{figure}

$F$ is a $q$-dependent linear mapping that raises the grade of its argument by one (e.g., it maps vectors to bivectors, etc. --- see Fig.~\ref{fig:FieldStr}).
It satisfies
\begin{equation} \label{FieldStrMulti}
F(A_r \wedge B_s) = F(A_r) \wedge B_s + (-1)^r A_r \wedge F(B_s)
\end{equation}
for any $r$-vector $A_r$ and $s$-vector $B_s$. Iterating this identity (see Eq.~(\ref{FieldStrMulti2}) in Appendix~\ref{sec:GAGCLinFunc}), together with the fact that $F(\alpha)=0$ for scalars $\alpha$, we find that $F$ is completely determined by its action on vectors.

We call $\oline{h}$ a \emph{pure gauge}, if $\oline{h}=\oline{f}$ for some diffeomorphism $f$. In this case
\begin{equation}
F(b) = -\oline{f} (\partial_q \wedge \partial_q) f^{-1} \cdot b = 0
\end{equation}
for all constant vectors $b$, i.e., the corresponding field strength vanishes. Vice versa, if $F(b)=0$, i.e.,
\begin{equation}
\partial_q \wedge \oline{h^{-1}} (b) = 0 , 
\end{equation}
then vector field $\oline{h^{-1}}(b;q)$ has scalar potential $\phi(b;q)$, and hence can be expressed as
\begin{equation}
\oline{h^{-1}}(b;q) = \partial_q \phi(b;q) = \partial_q f^{-1}(q) \cdot b ,
\end{equation}
where $f^{-1}$ is a vector representing the linear functional $\phi(b)$.

For a generic gauge field it is easy to show (see Eq.~(\ref{FieldStrTrPf}) in Appendix~\ref{sec:GAGCLinFunc}) that expression $F(\bar{P})$ is invariant under gauge transformations $f$,
\begin{equation}
F'(\bar{P}') = F(P) .
\end{equation}
This is indeed a crucial property expected of a gauge field strength that legitimizes its definition in Eq.~(\ref{FieldStrength}), and also confirms gauge invariance of the canonical equations~(\ref{CanEOMgauge}).

The here-defined gauge field strength $F$ aims to be a universal concept unifying gauge field strengths of Yang-Mills and gravity theories. Its relation to the traditional Yang-Mills field strength is provided by Eq.~(\ref{FieldStrYMTrad}), and its relevance for gravity stressed in Appendix~\ref{sec:FieldStrTorsion}, where it is interpreted as the torsion of the teleparallel theory of gravity.

\section{Symmetries of gauged Hamiltonian}
\label{sec:SymGaugedHam}

The symmetry condition~(\ref{SymCrit}) for a gauged Hamiltonian $H_h$, expressed in terms of the original Hamiltonian $H$, reads 
\begin{equation}
H\big(f(q),\oline{h}(\oline{f^{-1}}(P;q);f(q))\big) = H(q,\oline{h}(P;q)) .
\end{equation}
Here, $f$ may be any transformation of the configuration space, in particular, we do not assume, at this stage, that it obeys Eq.~(\ref{GlobSym}).

The infinitesimal version of the symmetry condition, Eq.~(\ref{SymCritInfsm}), for Hamiltonian $H_h$ can be expanded in terms of $H$ using Eqs.~(\ref{GaugedHamDifP}) and (\ref{GaugedHamDifq}),
\begin{equation} \label{SymInfsmGauge}
v \cdot \dot{\partial}_q H(\dot{q},\bar{P})
+ \left[ v \cdot \dot{\partial}_q \dot{\oline{h}}(P)
- \oline{h}\big( \dot{\partial}_q \wedge (\dot{v} \cdot P) \big) \right]
\cdot \partial_{\bar{P}} H(q,\bar{P})
= 0 .
\end{equation}
If we now assume that $f(q) = q + \eps\, v(q)$ is such that Eq.~(\ref{GlobSym}) holds, the first term vanishes, and we observe that for $f$ to be a symmetry of $H_h$ it is sufficient that
\begin{equation} \label{SymGenUniv}
v \cdot \dot{\partial}_q \dot{\oline{h}}(P)
= \oline{h}\big( \dot{\partial}_q \wedge (\dot{v} \cdot P) \big) ,
\end{equation}
regardless of a concrete form of $H$. This condition, however, may be too restrictive --- a vector field $v$ may well be a symmetry generator according to Eq.~(\ref{SymInfsmGauge}) for a given Hamiltonian $H$, even though it fails to satisfy Eq.~(\ref{SymGenUniv}). 
In Example~\ref{sec:ExStrGr}, we will show how Eq.~(\ref{SymInfsmGauge}) reduces to the Killing equation of general relativity, and identify $v$ with the Killing vector.

The conservation law corresponding to a symmetry generated by a vector field $v$ maintains its usual form (recall Eq.~(\ref{PreConsLaw}))
\begin{align} \label{ConsLaw}
d\Gamma \cdot \partial_q \, (P \cdot v)
&= 0  \hspace{10mm}{\rm for}~ D=1  \nonumber\\
(d\Gamma \cdot \partial_q) \cdot (P \cdot v)
&= 0 \hspace{10mm}{\rm for}~ D>1 .
\end{align}
However, note that the relation between $P$ and $d\Gamma$, which is deduced from the canonical equations~(\ref{CanEOM1gauge}) and (\ref{CanEOM3gauge}), is altered due to the presence of the gauge field $\oline{h}$ in the gauged Hamiltonian $H_h$.

\section{Examples}
\label{sec:Examples}

In Examples \ref{sec:ExGr}, \ref{sec:ExYM} and \ref{sec:ExSF}, the configurations space $\mathcal{C}$ is understood as a Cartesian product of $D$-dimensional spacetime (the ``$x$-space" with pseudoscalar $I_x$), and $N$-dimensional space of fields (the ``$y$-space" with pseudoscalar $I_y$). The points in $\mathcal{C}$ are decomposed accordingly as $q = x + y$, and vectors as $a = a_x + a_y$, where $a_x \equiv a \cdot I_x I_x^{-1}$ and $a_y \equiv a \cdot I_y I_y^{-1}$ are respective spacetime and field-space projections. The vector derivative operator is likewise decomposed as $\partial_q = \partial_x + \partial_y$. Whenever convenient, we will introduce a spacetime basis $\{\gamma_\mu\}_{\mu=1}^D$ and its reciprocal $\{\gamma^\mu\}_{\mu=1}^D$ (see Appendix~\ref{sec:GAGCBasis}), as well as an orthonormal  basis $\{e_a\}_{a=1}^N$ of the field space, and assume Einstein summation convention over repeated indices. We leave the signature of spacetime arbitrary, however, the signature of the field space is, for simplicity, assumed to be Euclidean.

The first two examples do not presume any particular form of the Hamiltonian. There, we are only concerned with certain classes of transformations of the configuration space, and introduce complementary gauge fields with correct transformation properties, dictated by Eq.~(\ref{TrGaugeField}). In Example~\ref{sec:ExGr}, spacetime diffeomorphisms give rise to a gravitational gauge field, while in Example~\ref{sec:ExYM} we investigate local (i.e., spacetime-variable) rotations in the field space, and introduce a Yang-Mills gauge field. Results of these two examples are applied in Example~\ref{sec:ExSF} where we choose the Hamiltonian of a scalar field theory.

The last Example~\ref{sec:ExStrGr} is independent of the previous ones. It studies relativistic particles or strings in a spacetime endowed with a generic gauge field.

\subsection{Spacetime diffeomorphisms and gravitational gauge field}
\label{sec:ExGr}

Gravitational field arises from the requirement of invariance under arbitrary spacetime diffeomorphisms $f_x(x)$. Let us therefore consider transformations of the configuration space $\mathcal{C}$ of the form
\begin{equation} \label{SpacetimeDiffeo}
q' = f_{\rm Gr}(q) = f_x(x) + y ,
\end{equation}
whose adjoint mapping
\begin{equation}
\oline{f}_{\rm Gr}(b;q) = \oline{f}_x(b_x;x) + b_y
\end{equation}
is determined by an $x$-dependent linear function $\oline{f}_x(b_x) = \partial_x f_x(x) \cdot b_x$, which maps spacetime vectors to spacetime vectors.

To satisfy Eq.~(\ref{TrGaugeField}) it is sufficient to assume that the corresponding gauge field is of the form
\begin{subequations} \label{GaugeFieldGr}
\begin{equation}
\oline{h}_{\rm Gr}(b;q)
= \oline{h}_x(b_x;x) + b_y ,
\end{equation}
where $\oline{h}_x$ is the restriction of $\oline{h}_{\rm Gr}$ to the $x$-space, whose transformation rule is
\begin{equation}
\oline{h'_x}(b_x;x') = \oline{h}_x\big( \oline{f}_x(b_x;x) ;x\big) .
\end{equation}
Corresponding derived forms of the gauge field are
\begin{align}
h_{\rm Gr}^{-1}(a) &= h_x^{-1}(a_x) + a_y , \nonumber\\
\oline{h^{-1}_{\rm Gr}}(b) &= \oline{h_x^{-1}}(b_x) + b_y , \nonumber\\
h_{\rm Gr}(a) &= h_x(a_x) + a_y .
\end{align}
\end{subequations}

The field strength of the gravitational gauge field, as obtained from Eq.~(\ref{FieldStrength}), reads
\begin{equation}
F_{\rm Gr}(b)
= -\oline{h}_{\rm Gr}\big(\dot{\partial}_q \wedge \dot{\oline{h^{-1}_{\rm Gr}}}(b)\big)
= -\oline{h}_{x}\big(\dot{\partial}_x \wedge \dot{\oline{h^{-1}_{x}}}(b_x)\big) .
\end{equation}
We may perceive that it annihilates field-space vectors, 
\begin{equation}
F_{\rm Gr}(b_y) = 0 ,
\end{equation}
and maps spacetime vectors to spacetime bivectors. Hence, 
\begin{equation} \label{FGrMulti1}
F_{\rm Gr}(I_x) = 0 ,
\end{equation}
since a $D+1$-vector in a $D$-dimensional spacetime necessarily vanishes; and, by virtue of Eq.~(\ref{FieldStrMulti}), we find
\begin{equation} \label{FGrMulti2}
F_{\rm Gr}\big(e_a \wedge (\gamma_\mu \cdot I^{-1}_x)\big) 
= -e_a \wedge F_{\rm Gr}(\gamma_\mu \cdot I^{-1}_x) ,
\end{equation}
where
\begin{equation}
F_{\rm Gr}(\gamma_\mu \cdot I^{-1}_x)
= -\oline{h}_x \big[ \dot{\partial}_x \wedge \big( \dot{h}_x(\gamma_\mu) \cdot \dot{\oline{h_x^{-1}}}(I_x^{-1}) \big) \big]
= - \gamma_\mu \cdot \dot{\oline{h}}_x(\dot{\partial}_x) \, \oline{h}_x \dot{\oline{h_x^{-1}}}(I_x^{-1}) 
\end{equation}
is a scalar multiple of $I_x$.

\subsection{Local field-space rotations and Yang-Mills gauge field}
\label{sec:ExYM}

In Yang-Mills theory with the rotation gauge group $SO(N)$, the Yang-Mills gauge field is introduced to impose invariance under local field-space rotations of the form
\begin{equation} \label{YMRot}
q' = f_{\rm YM}(q) = R(x) q \rev{R}(x),\quad R(x) = e^{-B_y(x)/2} ,
\end{equation}
where $B_y$ is an $x$-dependent field-space bivector, i.e., $B_y \cdot I_y = B_y I_y$. (See Appendix~\ref{sec:GAGCRot} for details on geometric algebra representation of rotations).

For an arbitrary constant vector $b$, the induced adjoint mapping is calculated
\begin{equation}
\oline{f}_{\rm YM}(b;q)
= \dot{\partial}_q (R \dot{q} \rev{R}) \cdot b + \dot{\partial}_q (\dot{R} q \dot{\rev{R}}) \cdot b
= \rev{R} b R + \dot{\partial}_x \big( y' \cdot (2 R \dot{\rev{R}}) \big) \cdot b ,
\end{equation}
where we have used Eq.~(\ref{RotorDif2}).
From the structure of $\oline{f}_{\rm YM}$ (namely, we know from Eq.~(\ref{RotorDif}) that $R \dot{\rev{R}}$ is a bivector) we infer a possible form of the corresponding Yang-Mills gauge field:
\begin{equation} \label{YMGaugeField}
\oline{h}_{\rm YM}(b;q)
= \rev{S}(x) b S(x) - \gamma^\mu \big( y \cdot A_\mu(x) \big) \cdot b ,
\end{equation}
where $S$ is a field-space rotor (just like $R$), and $\{A_\mu\}_{\mu=1}^D$ a set of field-space bivectors. (For simplicity, we have not introduced the gauge coupling constant.)

In order to find the transformation rules for $A_\mu$ and $S$, we compose, according to Eq.~(\ref{TrGaugeField}),
\begin{align}
\oline{h}_{\rm YM}\big(\oline{f}_{\rm YM}(b;q);q \big)
&= \rev{S} \rev{R} b R S + \dot{\partial}_x \big( y' \cdot (2 R \dot{\rev{R}}) \big) \cdot b - \gamma^\mu \big( y \cdot A_\mu(x) \big) \cdot (\rev{R} b R) \nonumber\\
&= \rev{RS} b RS + \gamma^\mu \big[ y' \cdot (2 R \partial_\mu \rev{R}) - y' \cdot (R A_\mu \rev{R}) \big] \cdot b .
\end{align}
Comparison with
\begin{equation}
\oline{h'}_{\rm YM}(b;q')
= \rev{S}'(x) b S'(x) - \gamma^\mu \big( y' \cdot A'_\mu(x) \big) \cdot b
\end{equation}
then yields
\begin{equation} \label{TrRulesYM}
S' = R S \quad,\quad A'_\mu = R A_\mu \rev{R} - 2 R \partial_\mu \rev{R} ,
\end{equation}
where we have denoted $\partial_\mu \equiv \gamma_\mu \cdot \partial_x$.

Since $\rev{S} \gamma^\mu S = \gamma^\mu$, we may cast Eq.~(\ref{YMGaugeField}) as
\begin{subequations} \label{GaugeFieldYM}
\begin{equation}
\oline{h}_{\rm YM}(b)
= \rev{S}\, \oline{h}_{\rm YM^*}(b) \,S \quad,\quad
\oline{h}_{\rm YM^*}(b) 
\equiv b - \gamma^\mu \big( y \cdot A_\mu \big) \cdot b .
\end{equation}
Note that $\oline{h}_{\rm YM^*}$ has the structure of a \emph{shear} linear mapping discussed in Appendix~\ref{sec:GAGCShear}. This observation allows us to easily find other derived forms of the gauge field:
\begin{align}
h^{-1}_{\rm YM}(a)
&= \rev{S}\, h^{-1}_{\rm YM^*}(a) \,S \quad,\quad
h^{-1}_{\rm YM^*}(a) = a + a \cdot \gamma^\mu \, y \cdot A_\mu  \nonumber\\
\oline{h}^{\,-1}_{\rm YM}(b)
&= \oline{h}^{\,-1}_{\rm YM^*}(S b \rev{S}) \quad,\quad
\oline{h}^{\,-1}_{\rm YM^*}(b) = b + \gamma^\mu ( y \cdot A_\mu) \cdot b  \nonumber\\
h_{\rm YM}(a)
&= h_{\rm YM^*}(S a \rev{S}) \quad,\quad
h_{\rm YM^*}(a) = a - a \cdot \gamma^\mu \,  y \cdot A_\mu .
\end{align}
\end{subequations}

The Yang-Mills field strength $F_{\rm YM}$ is obtained from the defining Eq.~(\ref{FieldStrength}) as follows. First, for a constant vector $b$ we calculate
\begin{align}
\partial_q \wedge \oline{h}^{\,-1}_{\rm YM}(b) 
&= \dot{\partial}_q \wedge \dot{\oline{h}^{\,-1}_{\rm YM^*}}(S b \rev{S})
+ \dot{\partial}_q \wedge \oline{h}^{\,-1}_{\rm YM^*}(\dot{S} b \dot{\rev{S}}) \nonumber\\
&= \dot{\partial}_x \wedge \gamma^\mu (y \cdot \dot{A}_\mu) \cdot (S b \rev{S})
+ \big( A_\mu \cdot (S b \rev{S}) \big) \wedge \gamma^\mu
+ \dot{\partial}_x \wedge \oline{h}^{\,-1}_{\rm YM^*}\big( (2 \dot{S} \rev{S})\cdot(S b \rev{S}) \big) ,
\end{align}
where we have used Eq.~(\ref{RotorDif2}). Now, since $\oline{h}^{\,-1}_{\rm YM}(b_x)=b_x$ for any spacetime vector $b_x$, we find
\begin{align}
F_{\rm YM}(b) 
&= - \oline{h}_{\rm YM} \big( \partial_q \wedge \oline{h}^{\,-1}_{\rm YM}(b) \big) \nonumber\\
&= - \dot{\partial}_x \wedge \gamma^\mu (y \cdot \dot{A}_\mu) \cdot (S b \rev{S})
- \oline{h}_{\rm YM} \big( A_\mu \cdot (S b \rev{S}) \big) \wedge \gamma^\mu
- \dot{\partial}_x \wedge \big( (2 \rev{S} \dot{S})\cdot b \big) \nonumber\\
&= \gamma^\mu \wedge \gamma^\nu \left[ y \cdot (\partial_\nu A_\mu) - (y \cdot A_\nu) \cdot A_\mu \right] \cdot (S b \rev{S})
+ \gamma^\mu \wedge \big( ( \rev{S} A_\mu S - 2 \rev{S} \partial_\mu S ) \cdot b \big) .
\end{align}
Notice that 
\begin{equation} \label{FieldStrYMx}
F_{\rm YM}(b_x) = 0 ,
\end{equation}
and that the pure spacetime component of the bivector $F_{\rm YM}(b)$, to which only the first term contributes, reads
\begin{equation} \label{FieldStrYMProjX}
(\gamma_\rho \wedge \gamma_\sigma) \cdot F_{\rm YM}(b)
= \big( y \cdot (\partial_\rho A_\sigma - \partial_\sigma A_\rho - A_\rho \times A_\sigma) \big) \cdot (S b \rev{S})
\end{equation}
where we have used the identity $(\gamma_\rho \wedge \gamma_\sigma) \cdot (\gamma^\mu \wedge \gamma^\nu) = \delta_\sigma^\mu \delta_\rho^\nu - \delta_\sigma^\nu \delta_\rho^\mu$, and Formula~(\ref{Jacobi2}).

Action of the field strength on arbitrary multivectors can be inferred from Eq.~(\ref{FieldStrMulti}). In particular, due to the property~(\ref{FieldStrYMx}),
\begin{equation} \label{FYMMulti1}
F_{\rm YM}(I_x) = 0 ,
\end{equation}
and
\begin{equation} \label{FYMMulti2}
F_{\rm YM}\big(e_a \wedge (\gamma_\mu \cdot I_x^{-1})\big)
= F_{\rm YM}\big(e_a) \wedge (\gamma_\mu \cdot I_x^{-1})
= - ( \rev{S} A_\mu S - 2 \rev{S} \partial_\mu S ) \cdot e_a \, I_x^{-1} ,
\end{equation}
where we have used the fact that $\gamma^\rho \wedge \gamma^\nu \wedge (\gamma_\mu \cdot I_x^{-1}) = 0$, and $\gamma^\nu \wedge (\gamma_\mu \cdot I_x^{-1}) = \delta_\mu^\nu I_x^{-1}$.

\subsection{Scalar field coupled to a gauge field}
\label{sec:ExSF}

It has been discussed in our previous article \cite{ZatlSymConsLaws} that an $N$-component scalar field can be described in the Hamiltonian constraint formalism by the Hamiltonian
\begin{equation} \label{HamSF}
H_{{\rm SF}}(q,P) = P \cdot I_x + \frac{1}{2} \sum_{a=1}^N \big( I_x \cdot (P \cdot e_a) \big)^2 + V(y) ,
\end{equation}
where $\{ e_a \}_{a=1}^N$ is an arbitrary orthonormal basis of the space of fields. As shown in Sec.~\ref{sec:StatGaugeField}, coupling to a gauge field $h$ is achieved simply by the replacement $P \rightarrow \oline{h}(P) \equiv \bar{P}$. The ensuing gauged scalar field Hamiltonian reads
\begin{equation} \label{HamSFh}
H_{{\rm SF},h}(q,P) = \bar{P} \cdot I_x + \frac{1}{2} \sum_{a=1}^N \big( I_x \cdot (\bar{P} \cdot e_a) \big)^2 + V(y) .
\end{equation}

The first canonical equation~(\ref{CanEOM2gauge}) takes the form
\begin{equation} \label{CanEOM1SFh}
A \cdot h^{-1}(d\Gamma)
= \lambda A \cdot \partial_{\bar{P}} H_{{\rm SF},h}
= \lambda A \cdot I_x + \lambda \sum_{a=1}^N \big( I_x \cdot (A \cdot e_a) \big) \cdot \big( I_x \cdot (\bar{P} \cdot e_a) \big)  ,
\end{equation}
where $A$ is an arbitrary $D$-vector.
It can be used, together with the relation $P \cdot d\Gamma = \bar{P} \cdot h^{-1}(d\Gamma)$, and the Hamiltonian constraint $H_{{\rm SF},h}(q,P) = 0$, to cast the extended action~(\ref{ActionAugm}) corresponding to the Hamiltonian $H_{{\rm SF},h}$ as
\begin{equation}
\mathcal{A}_{{\rm SF},h}[\gamma,P,\lambda]
= \int_\gamma \lambda \left[ \bar{P} \cdot \partial_{\bar{P}} H_{{\rm SF},h} - H_{{\rm SF},h} \right]
= \int_\gamma \lambda \left[ \frac{1}{2} \sum_{a=1}^N \big( I_x \cdot (\bar{P} \cdot e_a) \big)^2 - V(y) \right] .
\end{equation}
The choice $A = I_x^{-1}$ in Eq.~(\ref{CanEOM1SFh}) identifies the Lagrange multiplier $\lambda$ as
\begin{equation} \label{LambdaSF}
\lambda = I_x^{-1} \cdot h^{-1}(d\Gamma) .
\end{equation}
From the same equation we also infer that
\begin{equation} \label{MomSF}
I_x^{-1} \cdot \big(h^{-1}(d\Gamma) \cdot e_a \big)
= \lambda I_x \cdot (\bar{P} \cdot e_a) .
\end{equation}
These relations allow us to write the action as a functional of the surface $\gamma$ alone,
\begin{equation} \label{ActionSFh2}
\mathcal{A}_{{\rm SF},h}[\gamma]
= \int_\gamma \left[ \frac{1}{2 \lambda} \sum_{a=1}^N \big( I_x^{-1} \cdot (h^{-1}(d\Gamma) \cdot e_a) \big)^2 - \lambda V(y) \right] .
\end{equation}

The equation of motion for $\gamma$ can be found either by varying this action, or from the canonical equations of motion~(\ref{CanEOMgauge}) with a help of relations (\ref{FGrMulti1}) and (\ref{FGrMulti2}) for the gravitational field, or (\ref{FYMMulti1}) and (\ref{FYMMulti2}) for the Yang-Mills field. Nevertheless, in the remaining part of this example we shall stay with our considerations on the level of action, where calculations are significantly less involved.

\subsubsection{Gravity}

The scalar field Hamiltonian~(\ref{HamSF}) is independent of $x$ and therefore satisfies condition~(\ref{GlobSym}) for spacetime diffeomorphisms~(\ref{SpacetimeDiffeo}).

Representing the motions as
$\gamma = \{ x + y(x) \,|\, x \in \Omega \}$, 
where $\Omega$ is a spacetime region, 
we find for the surface element (see Ref.~\cite{ZatlSymConsLaws})
\begin{equation} \label{dGamma}
d\Gamma = dX + (dX \cdot \partial_x) \wedge y + \ldots ,
\end{equation}
where ``$\ldots$" gathers terms with two and more $y$-space components. Here, $dX=|dX| I_x$ is the oriented infinitesimal element of the spacetime with magnitude $|dX|$ (this is the usual scalar Riemann measure, traditionally denoted by $d^Dx$), and the orientation defined by the spacetime pseudoscalar $I_x$.

Let us consider the gravitational gauge field $h_{\rm Gr}$ characterized by Eqs.~(\ref{GaugeFieldGr}). Since it acts on $y$-space vectors as an identity, $h_{\rm Gr}(a_y)=a_y$, we find for the Lagrange multiplier given by Eq.~(\ref{LambdaSF})
\begin{equation} \label{LambdaGr}
\lambda_{\rm Gr} = I_x^{-1} \cdot h_{\rm Gr}^{-1}(dX)
= |dX| \det(h_{\rm Gr}^{-1}) ,
\end{equation}
where the determinant is defined in geometric algebra terms in Appendix~\ref{sec:GAGCLinFunc}. Moreover, we have
\begin{equation}
I_x^{-1} \cdot (h_{\rm Gr}^{-1}(d\Gamma) \cdot e_a) 
= I_x^{-1} \cdot \big( h_{\rm Gr}^{-1}(dX) \cdot \oline{h}_{\rm Gr}(\partial_x \phi_a) \big)
= |dX| \det(h_{\rm Gr}^{-1}) \, \oline{h}_{\rm Gr}(\partial_x \phi_a) ,
\end{equation}
where the scalars $\phi_a \equiv e_a \cdot y$ are the components of $y$.

The action~(\ref{ActionSFh2}) now reads
\begin{equation} \label{ActionGr}
\mathcal{A}_{{\rm SF,Gr}}[y(x)]
= \int_\Omega \det(h_{\rm Gr}^{-1}) \left[ \frac{1}{2} \sum_{a=1}^N \big( \oline{h}_{\rm Gr}(\partial_x \phi_a) \big)^2 - V(y) \right] |dX| ,
\end{equation}
which can be further elucidated by writing
\begin{equation}
\sum_{a=1}^N \big( \oline{h}_{\rm Gr}(\partial_x \phi_a) \big)^2
= (\partial_x \phi_a) \cdot h_{\rm Gr} \oline{h}_{\rm Gr}(\partial_x \phi_a)
= g^{\mu\nu} \partial_\mu \phi_a \partial_\nu \phi_a ,
\end{equation}
where
\begin{equation}
g_{\mu\nu} = \gamma_\mu \cdot \oline{h_{\rm Gr}^{-1}} h_{\rm Gr}^{-1}(\gamma_\nu) \quad,\quad
g^{\mu\nu} = \gamma^\mu \cdot h_{\rm Gr} \oline{h}_{\rm Gr}(\gamma^\nu)
\end{equation}
are the components of the metric tensor of general relativity, and its inverse, respectively. 

The metric, regarded as a linear mapping $g=\oline{h_{\rm Gr}^{-1}} h_{\rm Gr}^{-1}$, has determinant
\begin{equation}
\det(g) = \det(h_{\rm Gr}^{-1})^2 .
\end{equation}
At the same time, by Formula~(\ref{DetExpand}),
\begin{equation}
\det(g) = I_x^{-1} \cdot g(I_x)
= \det(\gamma^\mu \cdot g(\gamma_\nu))
= \det(\gamma^\lambda \cdot \gamma^\kappa) \det(g_{\mu\nu})
= -\det(g_{\mu\nu}) ,
\end{equation}
where in the last equality we assumed, for definiteness, that the $x$-space has Lorentzian signature $\eta=(1,-1,-1,-1)$, and $\gamma^\mu \cdot \gamma^\nu=\eta^{\mu\nu}$.

These observations should be enough to conclude that Eq.~(\ref{ActionGr}) depicts the action of a scalar field coupled to a gravitational field (c.f. \cite[Ch.~6.4]{Ramond}), with $h_{\rm Gr}$ playing the role of vierbein of tetrad gravity formulations. We shall have more to say about the relation between the present gauge-field approach and the standard metric formulation of general relativity in Example~\ref{sec:ExStrGr}.

\subsubsection{Yang-Mills field}

To ensure that the scalar field Hamiltonian~(\ref{HamSF}) satisfies condition~(\ref{GlobSym}) for field-space rotations~(\ref{YMRot}), we will now assume that the potential is of the form
\begin{equation}
V(y) = U(y^2) .
\end{equation}

For the Yang-Mills gauge field $h_{\rm YM}$ characterized by Eqs.~(\ref{GaugeFieldYM}), and the surface element of $\gamma$ given by Eq.~(\ref{dGamma}), Eq.~(\ref{LambdaSF}) reads
\begin{equation} \label{LambdaYM}
\lambda_{\rm YM} = I_x^{-1} \cdot h_{\rm YM}^{-1}(d\Gamma)
= I_x^{-1} \cdot dX = |dX| .
\end{equation}
Here we have used relations $S I_x^{-1} \rev{S} = I_x^{-1}$, and $h^{-1}_{\rm YM^*}(a_y) = a_y$.
Furthermore, since
\begin{equation}
h^{-1}_{\rm YM^*}(dX) = dX + (dX \cdot \gamma^\mu) \wedge (y \cdot A_\mu) + \ldots ,
\end{equation}
where ``$\ldots$" gathers terms with two and more $y$-space components, 
\begin{align} \label{KinScFYM}
I_x^{-1} \cdot \big(h_{\rm YM}^{-1}(d\Gamma) \cdot e_a \big) 
&= I_x^{-1} \cdot \big(h_{\rm YM^*}^{-1}(d\Gamma) \cdot (S e_a \rev{S}) \big) \nonumber\\
&= I_x^{-1} \cdot \big[ h_{\rm YM^*}^{-1}(dX) \cdot (S e_a \rev{S})
+ \big( h_{\rm YM^*}^{-1}(dX \cdot \partial_x) \wedge y \big) \cdot (S e_a \rev{S}) \big] \nonumber\\
&= |dX| \big[\gamma^\mu (y \cdot A_\mu)\cdot (S e_a \rev{S}) + \dot{\partial}_x \dot{y} \cdot (S e_a \rev{S}) \big] \nonumber\\
&= |dX| \gamma^\mu (\partial_\mu y + y \cdot A_\mu ) \cdot (S e_a \rev{S}) .
\end{align}
Substituting Eqs.~(\ref{LambdaYM}) and (\ref{KinScFYM}) into Eq.~(\ref{ActionSFh2}), we find the action of a scalar field coupled to a Yang-Mills gauge field
\begin{equation} \label{ActionSFYM}
\mathcal{A}_{\rm SF,YM}[y(x)] 
= \int_\Omega \left[ \frac{1}{2} (\partial_\mu y + y \cdot A_\mu) \cdot (\partial^\mu y + y \cdot A^\mu) - U(y^2) \right] |dX| .
\end{equation}
This action is independent of $S$ due to the invariance of the scalar field Hamiltonian~(\ref{HamSF}) under global field-space rotations. 

Let us note that although we consider only real-valued fields, the present formalism allows us to discuss also (special) unitary groups of transformations, which play the most prominent role in physics. Indeed, any Lie algebra $\mathfrak{u}(n)$ can be represented as a bivector subalgebra of the geometric algebra of a $2n$-dimensional Euclidean space, while the Lie group $U(n)$ is represented by the corresponding group of rotors (see Refs. \cite[Ch.~11.4]{DoranLas} or \cite{Doran1993}).

In the kinetic part of action~(\ref{ActionSFYM}) we identify the traditional Yang-Mills covariant derivative
\begin{equation} \label{YMCovDer}
D_\mu y \equiv \partial_\mu y + y \cdot A_\mu = \partial_\mu y + \mathsf{A}_\mu(y) ,
\end{equation}
where $\mathsf{A}_\mu$ is a linear antisymmetric mapping characterized by the bivector $A_\mu$ (see Appendix~\ref{sec:GAGCBivectAntisym}).
Commutator of two covariant derivatives yields the traditional Yang-Mills field strength
\begin{equation}
\mathsf{F}_{\mu\nu}(y) \equiv (D_\mu D_\nu - D_\nu D_\mu) y 
= y \cdot (\partial_\mu A_\nu - \partial_\nu A_\mu - A_\mu \times A_\nu) .
\end{equation}
The last term on the right-hand side has been obtained by the Jacobi identity, Eq.~(\ref{Jacobi2}). For every $\mu$ and $\nu$, $\mathsf{F}_{\mu\nu}$ is an antisymmetric linear function acting on the $y$-space, with characteristic bivector
\begin{equation}
F_{\mu\nu} = \partial_\mu A_\nu - \partial_\nu A_\mu - A_\mu \times A_\nu .
\end{equation}
Eq.~(\ref{FieldStrYMProjX}) now establishes an explicit relation between the traditional field strength of an $SO(N)$ Yang-Mills theory and the Yang-Mills field strength defined according to Formula~(\ref{FieldStrength}):
\begin{equation} \label{FieldStrYMTrad}
(\gamma_\mu \wedge \gamma_\nu) \cdot F_{\rm YM}(b)
= \big( \mathsf{F}_{\mu\nu}(y) \big) \cdot (S b \rev{S}) .
\end{equation}

\subsubsection{Electromagnetic field}

Let us now specialize the Yang-Mills field to the electromagnetic field, i.e., take a two-dimensional field space ($N=2$) where all bivectors are scalar multiples of the pseudoscalar $I_y$. Due to this simplification, the rotor $R$ from Eq.~(\ref{YMRot}) can be parametrized by a single rotation angle $\theta$,
\begin{equation}
R_{\rm EM}(x) = e^{-I_y \theta(x) / 2} ,
\end{equation}
and the bivectors $A_\mu$ can be written as 
\begin{equation}
A_\mu = \alpha_\mu I_y ,
\end{equation}
where $\alpha_\mu$ are components of the electromagnetic vector four-potential (which is a spacetime vector). 

The electromagnetic gauge field has the form (c.f. Eq.~(\ref{YMGaugeField}))
\begin{equation}
\oline{h}_{\rm EM}(b;q) 
= e^{I_y \phi(x)/2} \,b\, e^{-I_y \phi(x)/2} - \gamma^\mu \alpha_\mu (y \cdot I_y) \cdot b ,
\end{equation}
and the transformation rules~(\ref{TrRulesYM}) reduce to
\begin{equation} \label{TrRulesEM}
\phi' = \phi + \theta \quad,\quad 
\alpha'_\mu = \alpha_\mu - \partial_\mu \theta ,
\end{equation}
as expected of the electromagnetic potential. The action~(\ref{ActionSFYM}) now reads
\begin{equation} \label{ActionSFEM}
\mathcal{A}_{\rm SF,EM}[y(x)] 
= \int_\Omega \left[ \frac{1}{2} (\partial_\mu y + \alpha_\mu y \cdot I_y) \cdot (\partial^\mu y + \alpha^\mu y \cdot I_y) - U(y^2) \right] |dX| .
\end{equation}

Scalar field that couples to electromagnetism is usually regarded as a complex field $\Phi(x) = \phi_R + i \phi_I$ with the action \cite{Peskin}
\begin{equation} \label{ActionEMComplex}
\mathcal{A}_{\rm SF,EM}[\Phi(x)] 
= \int_\Omega \left[ \frac{1}{2} (\mathcal{D}_\mu \Phi)^{*} \mathcal{D}^\mu \Phi - U(\Phi^{*} \Phi) \right] |dX| , 
\end{equation}
where $\mathcal{D}_\mu = \partial_\mu + i \alpha_\mu$ is the traditional covariant derivative (the coupling constant has been omitted). 
Our use of geometric algebra offers an equivalent formulation~(\ref{ActionSFEM}) in terms of a two-component vector
\begin{equation}
y = \phi_R e_R + \phi_I e_I = e_R \Phi ,
\end{equation}
where the imaginary unit of complex numbers $i$ has been identified with the unit pseudoscalar $I_y = e_R e_I$ of the two-dimensional field space (see Appendix~\ref{sec:GAGCComplex}). The correspondences
\begin{align}
\Phi^{*} \Phi 
&= (\phi_R - \phi_I e_R e_I) (\phi_R + \phi_I e_R e_I)
= y e_R e_R y = y^2 , \nonumber\\
\mathcal{D}_\mu \Phi 
&= (\partial_\mu + \alpha_\mu e_R e_I) (\phi_R + \phi_I e_R e_I)
= e_R (\partial_\mu y + \alpha_\mu y I_y) , \nonumber\\
(\mathcal{D}_\mu \Phi)^{*} 
&= (\partial_\mu - \alpha_\mu e_R e_I) (\phi_R - \phi_I e_R e_I)
= (\partial_\mu y + \alpha_\mu y I_y) e_R 
\end{align}
then show that the complex action~(\ref{ActionEMComplex}) is indeed equivalent with the real action~(\ref{ActionSFEM}).

\subsection{String coupled to gravity}
\label{sec:ExStrGr}

The Hamiltonian describing free relativistic particles, strings or high-dimensional membranes has been introduced in Example~5.3 of Ref.~\cite{ZatlCanEOM}. For spacetimes with arbitrary signature it reads
\begin{equation} \label{HamStr}
H_{\rm Str}(q,P) = \frac{1}{2} ( \rev{P} \cdot P - \Lambda^2 ) ,
\end{equation}
where $\Lambda$ is a positive constant, and ``$\rev{~~~~}$" is the reversion operation defined in Eq.~(\ref{GAGCRev}). (Positive $\rev{P} \cdot P$ can be replaced by $|P|^2$, where $|\,.\,|$ is the magnitude defined in Eq.~(\ref{GAGCMagnitude}).) Motions $\gamma$ are identified with world-sheets, and the configuration space $\mathcal{C}$ with the target space of string theory, i.e., with the spacetime. The corresponding gauged Hamiltonian
\begin{equation}
H_{{\rm Str},h}(q,P) = \frac{1}{2} \big( \oline{h}(\rev{P}) \cdot \oline{h}(P) - \Lambda^2 \big) 
\end{equation}
implements coupling to a static gravitational field described by the gauge field $h$. (For the case $D=1$ see also Ref.~\cite[Ch.~4(d)]{Lasenby1998}.)

The first canonical equation (\ref{CanEOM1gauge}) now reads
\begin{equation}
h^{-1}(d\Gamma) = \lambda \oline{h}(\rev{P}) \quad\Leftrightarrow\quad
P = \frac{1}{\lambda} \oline{h^{-1}} h^{-1}(\rev{d\Gamma}) .
\end{equation}
Taking the magnitude, and using the Hamiltonian constraint, Eq.~(\ref{CanEOM3gauge}), we find
\begin{equation}
|\lambda| \Lambda = |h^{-1}(d\Gamma)| .
\end{equation}
The latter two equations allow us to eliminate $P$ and $\lambda$, and cast the action~(\ref{ActionAugm}) as
\begin{equation}
\mathcal{A}_{{\rm Str},h}[\gamma]
= \int_\gamma P \cdot d\Gamma
= \int_\gamma \bar{P} \cdot h^{-1}(d\Gamma)
= \int_\gamma \frac{1}{\lambda} h^{-1}(\rev{d\Gamma}) \cdot h^{-1}(d\Gamma)
= \pm \Lambda \int_\gamma |h^{-1}(d\Gamma)| ,
\end{equation}
where ``$\pm$" is the sign of $\lambda$.
This action can be written solely in terms of the metric $g \equiv \oline{h^{-1}} h^{-1}$,
\begin{equation} \label{ActionStrGr}
\mathcal{A}_{{\rm Str},h}[\gamma]
= \pm \Lambda \int_\gamma \sqrt{\rev{d\Gamma} \cdot g(d\Gamma)} .
\end{equation}

As regards the remaining second canonical equation~(\ref{CanEOM2gauge}), one can show, using the identity~(\ref{GAhhDer}) and
\begin{equation} \label{StrGrRel1}
\dot{\partial}_q \dot{\oline{h^{-1}}} h^{-1}(U) \cdot U
= \frac{1}{2} \dot{\partial}_q \dot{\oline{h^{-1}}} \dot{h^{-1}}(U) \cdot U ,
\end{equation}
that it reads
\begin{equation} \label{GeodesicEqGA}
(-1)^{D-1} \frac{1}{2} \dot{\partial}_q \dot{g}(U) \cdot U
= \begin{cases}
U \cdot \partial_q g(U) & ~~{\rm for}~ D=1 \\
(U \cdot \partial_q) \cdot g(U) &  ~~{\rm for}~ D>1 .
\end{cases}
\end{equation}
Here we have introduced a generalization of the four-velocity $U \equiv d\Gamma/|h^{-1}(d\Gamma)|$, which is normalized so that $\rev{U} \cdot g(U) = 1$.

\subsubsection{Component form of the geodesic equation}

It is instructive to compare our results with the traditional ``component" (or tensor) approach to general relativity. For this purpose we choose a basis $\{\gamma_\mu\}_{\mu=1}^{D+N}$, and its reciprocal $\{\gamma^\mu\}_{\mu=1}^{D+N}$, and write the components of the metric and its inverse
\begin{align}
g_{\mu\nu} = \gamma_\mu \cdot g(\gamma_\nu) \quad,\quad g = \oline{h^{-1}} h^{-1} , \nonumber\\
g^{\mu\nu} = \gamma^\mu \cdot g^{-1}(\gamma^\nu) \quad,\quad g^{-1} = h \oline{h} .
\end{align}

For simplicity, let us concentrate on the case $D=1$, i.e., relativistic particle. With a parametrization of the trajectory $\gamma = \{ q(\tau) \,|\, \tau \in [\tau_i,\tau_f] \}$ the action~(\ref{ActionStrGr}) takes the familiar form
\begin{equation}
\mathcal{A}_{{\rm Str},h}[q(\tau)]
= \pm \Lambda \int_{\tau_i}^{\tau_f} \sqrt{g_{\mu\nu} \frac{dq^{\mu}}{d\tau} \frac{dq^{\nu}}{d\tau}} d\tau .
\end{equation}

Furthermore, the equation of motion~(\ref{GeodesicEqGA}) is an equivalent of the geodesic equation \cite{Einstein1914}. This can be seen by rearranging
\begin{equation}
U \cdot \partial_q U 
= g^{-1} \left[ \frac{1}{2} \dot{\partial}_q \dot{g}(U) \cdot U 
- U \cdot \dot{\partial}_q \dot{g}(U) \right] ,
\end{equation}
and introducing the components $U^\mu=\gamma^\mu \cdot U$, $\partial_\mu=\gamma_\mu \cdot \partial_q$,
\begin{equation}
U^\nu \partial_\nu U^\mu 
= g^{\mu\nu} \left[ \frac{1}{2} \partial_\nu g_{\lambda\kappa} 
- \partial_\kappa g_{\nu\lambda} \right] U^\lambda U^\kappa .
\end{equation}
Recall the definition of the \emph{Christoffel symbols}
\begin{equation} \label{Christoffels}
\Gamma^\mu_{~\lambda \kappa} 
= \frac{1}{2} g^{\mu\nu} \left[ \partial_\kappa g_{\nu\lambda} + \partial_\lambda g_{\nu\kappa} -\partial_\nu g_{\lambda\kappa} \right] ,
\end{equation}
which have the symmetry property $\Gamma^\mu_{~\lambda \kappa} = \Gamma^\mu_{~\kappa \lambda}$, and take the parametrization of $\gamma$ such that $U = dq/d\tau$. Then, we finally arrive at the standard component form of the geodesic equation
\begin{equation}
\frac{d^2 q^\mu}{d\tau^2}+ \Gamma^\mu_{~\lambda \kappa} \frac{d q^\lambda}{d\tau} \frac{d q^\kappa}{d\tau} = 0 .
\end{equation}

\subsubsection{Symmetries and Killing equation}

Infinitesimal symmetries of a gauged Hamiltonian are found from Eq.~(\ref{SymInfsmGauge}), which for the present Hamiltonian~(\ref{HamStr}) reads
\begin{equation}
\left[ v \cdot \dot{\partial}_q \dot{\oline{h}}(P)
- \oline{h}\big( \dot{\partial}_q \wedge (\dot{v} \cdot P) \big) \right]
\cdot \oline{h}(\rev{P})
= 0 ,
\end{equation}
and has to be satisfied for all $P$.
With a help of Eq.~(\ref{StrGrRel1}), this can be cast purely in terms of the metric,
\begin{equation}
\left[ \frac{1}{2} v \cdot \dot{\partial}_q \dot{g^{-1}}(P)
- g^{-1}\big( \dot{\partial}_q \wedge (\dot{v} \cdot P) \big) \right]
\cdot \rev{P}
= 0 .
\end{equation}
Let us consider again the case $D=1$, and put $P=g(a)$. Using the symmetricity of $g$, $\oline{g}=g$, and the relation $\dot{g} g^{-1} = - g \dot{g^{-1}}$, we find
\begin{equation}
\left[ \frac{1}{2} v \cdot \dot{\partial}_q \dot{g}(a)
+ \dot{\partial}_q \dot{v} \cdot g(a) \right]
\cdot a
= 0 .
\end{equation}
According to Eq.~(\ref{AntisymConds}), this, in turn, is equivalent with the equation
\begin{equation}
v \cdot \dot{\partial}_q \, a \cdot \dot{g}(b)
+ a \cdot \dot{\partial}_q \, b \cdot g(\dot{v})
+ b \cdot \dot{\partial}_q \, a \cdot g(\dot{v})
= 0
\end{equation}
being satisfied for all vectors $a$ and $b$. In components, i.e., for $v=v^\lambda\gamma_\lambda$, $a = \gamma_\mu$ and $b = \gamma_\nu$, this condition reads
\begin{equation} \label{StrGrKilling}
v^\lambda \partial_\lambda g_{\mu\nu}
+ g_{\nu\lambda} \partial_\mu v^\lambda
+ g_{\mu\lambda} \partial_\nu v^\lambda
= 0 ,
\end{equation}
which is an equivalent form of the \emph{Killing equation} \cite[Ch.~25.2]{MisnerThorneWheeler}
\begin{equation}
v_{\mu;\nu} + v_{\nu;\mu} = 0 ,
\end{equation}
with the covariant derivative $v^{\kappa}_{~;\nu}=\partial_\nu v^\kappa + \Gamma^\kappa_{~\nu \lambda} v^\lambda$. Indeed, by expanding
\begin{equation}
v_{\mu;\nu} + v_{\nu;\mu}
= g_{\mu \kappa} v^{\kappa}_{~;\nu} + g_{\nu \kappa} v^{\kappa}_{~;\mu}
= g_{\mu \kappa} (\partial_\nu v^\kappa + \Gamma^\kappa_{~\nu \lambda} v^\lambda)
+ g_{\nu \kappa} (\partial_\mu v^\kappa + \Gamma^\kappa_{~\mu \lambda} v^\lambda) ,
\end{equation}
and substituting from Eq.~(\ref{Christoffels}), we obtain Eq.~(\ref{StrGrKilling}).

The symmetry generator $v$ is traditionally referred to as the \emph{Killing vector}. To each Killing vector corresponds a conserved quantity
\begin{equation}
P \cdot v = \frac{1}{\lambda} g(d\Gamma) \cdot v
= \pm\Lambda U \cdot g(v)
= \pm\Lambda g_{\mu\nu} U^\mu v^\nu
\end{equation}
(c.f. Eq.~(25.5) in \cite{MisnerThorneWheeler}), which satisfies the conservation law~(\ref{ConsLaw}).

\section{Conclusion} 

We have enriched the Hamiltonian constraint formulation of classical field theories by introducing a gauge field --- a position-dependent linear mapping that ensures invariance of the theory under an extended group of local transformations of the configuration space via coupling to the momentum multivector.
Canonical equations of motion for the ensuing gauged Hamiltonian have been derived (Eqs.~(\ref{CanEOMgauge})), and symmetry conditions for a fixed gauge field have been discussed.

We have proposed a generic form of the gauge field strength, which is a $q$-dependent linear mapping given by Eq.~(\ref{FieldStrength}) that raises the grade of its argument by one. It can be interpreted as torsion of the Weitzenb\"{o}ck connection used in the teleparallel theory of gravity.

In principle, all diffeomorphisms of the space of partial observables are gauged if the gauge field is allowed to have the most generic form. However, in examples with an $N$-component scalar field, we restricted the group of transformations under which the theory has to be invariant, respectively, to spacetime diffeomorphism, and local field-space rotations in order to restrict the form of the gauge field. In the first case we generated gravitational, and in the second case Yang-Mills interaction. The generic gauge field can be therefore viewed, at least for this ``toy model" of scalar field theory, as a unified classical field \cite{Goenner,GoennerII}. 

In our attempt to reconcile gravity and Yang-Mills theory we view gravitational field as a field on flat spacetime rather than as a metric of a curved Riemannian manifold. The viability of such approach to gravity has been shown by the Gauge Theory Gravity \cite{Lasenby1998,Hestenes2005}, which indeed offered significant inspiration for this article.

Dynamics of the gauge field has not been addressed. However, the definition of the field strength and the brief investigation of symmetries and conservation laws for the gauged Hamiltonian are important prerequisites for future development of dynamical equations for the unified gauge field.

Finally, let us remark that although this article has been concerned only with classical field theory, one of the main motivations for introducing Hamiltonian methods in field theory is the desire to develop new quantization schemes (see, e.g., Ref.~\cite{Kanat1999}) that could compete with established Lagrangian formalism.

\subsection*{Acknowledgement}

The author received support from
Czech Science Foundation (GA\v{C}R), Grant GA14-07983S, and Deutsche Forschungsgemeinschaft (DFG), Grant KL 256/54-1. In addition, he would like to thank the organizers and participants of the workshop ``Rethinking Foundations of Physics 2016", held in Dorfgastein, Austria, for creating a stimulating environment, in which many of the ideas of this article have been cultivated.

\appendix
\section{Elements of geometric algebra and calculus} \label{sec:GAGC}

Let $V$ be an $n$-dimensional real vector space of possibly mixed (but non-degenerate) signature, and let $\mathcal{G}(V)$ denote its geometric algebra \cite{Hestenes,DoranLas}. (For concise introduction see also appendices of \cite{ZatlCanEOM,ZatlSymConsLaws}.) In this appendix we quote some key definitions and results employed in the main text.

\emph{Magnitude} of a grade-$r$ multivector $A$ is defined
\begin{equation} \label{GAGCMagnitude}
|A| = \sqrt{|\rev{A} \cdot A|} ,
\end{equation}
where $|\rev{A} \cdot A|$ is the absolute value of the scalar $\rev{A} \cdot A$, and
\begin{equation} \label{GAGCRev}
\rev{A}=(-1)^{r(r-1)/2}A
\end{equation}
is the \emph{reversion} of $A$, which satisfies, for any two multivectors, $\rev{AB}=\rev{B}\rev{A}$. If $|A|\neq 0$ then $A/|A|$ is normalized to unity. 

A normalized highest-grade element of $\mathcal{G}(V)$ is referred to as the \emph{unit pseudoscalar} $I$. For any vector $a \in V$, $a \cdot I = a I$.

\emph{Projection} of a vector $a$ onto a blade (a decomposable multivector) $A$ is $a \cdot A A^{-1}$. It vanishes for vectors perpendicular to $A$, and equals $a$ if $a \wedge A = 0$, i.e., if the vector is parallel with $A$.

If $a$ is a vector, and $A_r$ and $B_s$ multivectors of grades $r$ and $s$, respectively, then the following identity holds (the proof can be found in \cite[Ch.~1-1]{Hestenes}):
\begin{align} \label{GAident1}  
(-1)^r a \wedge (A_r \cdot B_s) + (A_r \cdot a) \cdot B_s = A_r \cdot (a \wedge B_s) ~~~~~ {\rm for}~ s \geq r > 1 .
\end{align}

The \emph{commutator product} is defined between any two multivectors by
\begin{equation} \label{GAGCComProd}
A \times B := \frac{1}{2}(A B - B A) .
\end{equation}
For any three multivectors, it satisfies the \emph{Jacobi identity}
\begin{equation} \label{Jacobi}
A \times (B \times C) + B \times (C \times A) + C \times (A \times B) = 0 .
\end{equation}
For a vector $y$ and a bivector $A$, the commutator product reduces to the inner product,
\begin{equation}
y \times A = y \cdot A = - A \cdot y .
\end{equation} 
The Jacobi identity for one vector $y$ and two bivectors $A_\mu$ and $A_\nu$ then reads
\begin{equation} \label{Jacobi2}
y \cdot (A_\mu \times A_\nu) 
= (y \cdot A_\mu) \cdot A_\nu - (y \cdot A_\nu) \cdot A_\mu ,
\end{equation}
where $A_\mu \times A_\nu$ is again a bivector, as can be easily proven (see Eq.~(1.62) in \cite[Ch.~1-1]{Hestenes}, or \cite[Ch.~4.1.3]{DoranLas}).

\subsection{Basis and reciprocal basis} \label{sec:GAGCBasis}

\emph{Basis} of the vector space $V$ is a set of $n$ vectors $\{e_1,\ldots,e_n\}$ that are linearly independent, i.e., satisfy $\blade{e}{n} \neq 0$. From these, a basis for the entire geometric algebra $\mathcal{G}(V)$ can be built by repeated use of the outer product \cite[Ch.~1-3]{Hestenes}. 

We do not assume the basis $\{e_j\}_{j=1}^n$ to be orthonormal, i.e., in general, $e_j \cdot e_k \neq \delta_{j k}$. (In mixed-signature spaces such basis does not even exist.) Therefore, in order to write the expansion of a vector $a \in V$
\begin{equation}
a = a \cdot e_j e^j = a \cdot e^j e_j ,
\end{equation}
we need to define the \emph{reciprocal basis} $\{e^1,\ldots,e^n\}$ by the requirement
\begin{equation}
e^j \cdot e_k = \delta^j_k \quad (\forall j,k=1,\ldots,n).
\end{equation}
The reciprocal basis can be explicitly constructed \cite[Ch.~4.3]{DoranLas}:
\begin{equation} \label{RecBasis}
e^j = (-1)^{j-1} e_1 \wedge \ldots \wedge \check{e}_j \wedge \ldots \wedge e_n E_n^{-1} ,
\end{equation}
where
\begin{equation} \label{BasisPseudoscalar}
E_n \equiv \blade{e}{n} ,
\end{equation}
and the check indicates, as usual, that the term is missing from the expression.


Consider, for example, the Minkowski space with a basis $\{\gamma_0,\gamma_1,\gamma_2,\gamma_3\}$, such that $\gamma_\mu \cdot \gamma_\nu = \eta_{\mu\nu}$, where $\eta = {\rm diag}(1,-1,-1,-1)$. The geometric algebra of this vector space coincides with the Dirac algebra of $\gamma$ matrices. The quantity (\ref{BasisPseudoscalar}), usually denoted $\gamma_5$, has the inverse ${\gamma_5^{-1} = - \gamma_3 \gamma_2 \gamma_1 \gamma_0}$. It can be verified that Eq.~(\ref{RecBasis}) yields the reciprocal basis
\begin{equation}
\gamma^0 = \gamma_0 \quad,\quad \gamma^1 = -\gamma_1 \quad,\quad \gamma^2 = -\gamma_2 \quad,\quad \gamma^3 = -\gamma_3 .
\end{equation}

\subsection{Complex numbers} \label{sec:GAGCComplex}

Consider a two-dimensional real vector space $V_2$ with positive signature, and its orthonormal basis $\{e_R,e_I\}$. In real geometric algebra, complex numbers are naturally identified with $\mathcal{G}^{+}(V_2)$, the even subalgebra of $\mathcal{G}(V_2)$ (see \cite[Ch.~4-7]{Hestenes} or \cite[Ch.~2.3.3]{DoranLas}). The latter contains multivectors of the form
\begin{equation}
\Phi = \phi_R + \phi_I I_y ,
\end{equation}
where $\phi_R,\phi_I$ are scalars (the real and the imaginary part of the corresponding complex number), and $I_y = e_R e_I$ is the unit pseudoscalar of $\mathcal{G}(V_2)$. 

$I_y$ serves the same purpose as the imaginary unit $i$. One can easily check that $I_y^2=-1$, and that acting from the right, it rotates any vector $y = \phi_R e_R + \phi_I e_I$ clockwise by $90^\circ$,
\begin{equation}
y I_y = -\phi_I e_R + \phi_R e_I .
\end{equation}
Moreover, vectors $y \in V_2$ are in one-to-one correspondence with elements of $\mathcal{G}^{+}(V_2)$,
\begin{equation}
y = \phi_R e_R + \phi_I e_I = e_R \Phi \quad\Leftrightarrow\quad
\Phi = e_R y .
\end{equation}

Complex units $e^{i \theta}$ implement clockwise rotation of a complex number $\Phi$ through an angle $\theta$ simply via multiplication, $\Phi' = e^{i\theta} \Phi$. The corresponding vectors are related analogously by
\begin{equation}
y' = e_R \Phi' = e_R e^{I_y \theta} \Phi = e_R e^{I_y \theta} e_R y
= e^{-I_y \theta} y = e^{-I_y \theta/2} y e^{-I_y \theta/2} .
\end{equation}
The final expression coincides with the geometric algebra prescription for the rotation through an angle $\theta$ in a plane defined by the bivector $I_y$, and immediately generalizes to vector spaces with dimension greater than two using rotors \cite[Ch.~4.2]{DoranLas}. Indeed, rotors can be thought of as higher-dimensional unit complex numbers.

\subsection{Transformations and induced mappings} \label{sec:GAGCTrInd}

Let $f:q\mapsto q'$ be an invertible smooth mapping (a diffeomorphism) on the space of partial observables $\mathcal{C}$.  For a vector $a$ in the tangent space of $\mathcal{C}$ we define the derivative of $f$ in direction $a$,
\begin{equation} \label{GAdifMap}
\uline{f}(a;q) 
\equiv a \cdot \partial_q f(q)
:= \lim_{\eps \rightarrow 0} \frac{f(q+\eps a) - f(q)}{\eps} .
\end{equation}
This gives rise to a $q$-dependent linear function, the \emph{differential} of $f$, that maps vectors $a$ at point $q$ to vectors $\uline{f}(a)$ at $q'$. 
The \emph{adjoint} of $\uline{f}$, denoted $\oline{f}$, is defined by
\begin{equation} \label{GCadjoint}
\oline{f}(b;q) := \partial_q f(q) \cdot b ,
\end{equation}
so that for any two vectors $a$ and $b$ it observes the identity 
\begin{equation} \label{GAadjointIdVect}
b \cdot \uline{f}(a) = \oline{f}(b) \cdot a .
\end{equation}
The adjoint maps vectors $b$ at a point $q'$ to vectors $\oline{f}(b)$ at $q$.

It is natural to extend the domain of $\uline{f}$ and $\oline{f}$ so that they may act on generic multivectors by demanding linearity and the \emph{outermorphism} property \cite[Ch. 3-1]{Hestenes}
\begin{equation}
\uline{f}(A \wedge B) = \uline{f}(A) \wedge \uline{f}(B) 
\quad,\quad
\oline{f}(A \wedge B) = \oline{f}(A) \wedge \oline{f}(B) .
\end{equation}
(For scalar arguments one defines $\underline{f}(\alpha) = \overline{f}(\alpha) = \alpha$). 
For an $r$-vector $A_r$ and an $s$-vector $B_s$, the following useful generalizations of Eq.~(\ref{GAadjointIdVect}) hold \cite[Ch. 3-1]{Hestenes}
\begin{align} \label{GAadjointIdent}
A_r \cdot \overline{f}(B_s) = \overline{f}[ \underline{f}(A_r) \cdot B_s ] ~~~~~ {\rm for}~ r \leq s , \nonumber\\
\underline{f}(A_r) \cdot B_s = \underline{f}[ A_r \cdot \overline{f}(B_s) ] ~~~~~ {\rm for}~ r \geq s .
\end{align}
We will refer to the differential $\uline{f}$ and the adjoint $\oline{f}$ collectively as \emph{induced} mappings, since they are induced by the diffeomorphism $f$.

Let us consider an arbitrary multivector-valued function $F$ on $\mathcal{C}$. The chain rule for differentiation,
\begin{equation} \label{GCchainRule}
a \cdot \partial_q F(f(q)) 
= \uline{f}(a) \cdot \partial_{q'} F(q') ,
\end{equation}
shows that the vector derivative operator transforms under $f$ as
\begin{equation} \label{GCTrDeriv}
\partial_{q'} = \oline{f^{-1}}(\partial_q) 
\end{equation}
(c.f. the transformation of momentum, Eq.~(\ref{TrRules})). We also see that
\begin{equation}
\uline{f^{-1}} = \uline{f}^{-1} ~~~{\rm and}~~~
\oline{f^{-1}} = \oline{f}^{-1} .
\end{equation} 

The induced mappings $\uline{f}$ and $\oline{f}$ are functions of $q$ which can be further differentiated. Commutativity of directional derivatives then leads to the identity 
\begin{equation} \label{GCadjointCurl}
\partial_q \wedge \oline{f}(A) = 0 ,
\end{equation}
valid for any constant multivector $A$.

\subsection{Linear functions} \label{sec:GAGCLinFunc}

Let $h:V \rightarrow V$ be a linear map. We can extend $h$ to an outermorphism $h:\mathcal{G}(V) \rightarrow \mathcal{G}(V)$, and introduce its adjoint $\oline{h}$ along the same lines as in the previous section for the induced mapping $\uline{f}$.

Outermorphism $h$ acting on a pseudoscalar $I$ produces a new pseudoscalar $h(I)$. The proportionality constant between the two pseudoscalars is the \emph{determinant} of $h$,
\begin{equation} \label{DetDef}
h(I) = I \det(h) \quad\Leftrightarrow\quad
\det(h) = I^{-1} h(I) = I^{-1} \cdot h(I) .
\end{equation}
Utilizing an arbitrary basis $\{e_j\}_{j=1}^n$ of $V$, and the reciprocal basis $\{ e^j \}_{j=1}^n$, it can be expressed as the determinant of the matrix $h^j_{~k} = e^j \cdot h(e_k)$ (see Formula~(4.12) in \cite[Ch.~1-4]{Hestenes}),
\begin{equation} \label{DetExpand}
\det(h) = (e^n \wedge \ldots \wedge e^1) \cdot \big(h(e_1) \wedge \ldots \wedge h(e_n) \big)
= \det(h^j_{~k}) .
\end{equation}
From definition~(\ref{DetDef}), two popular properties of determinants,
\begin{equation}
\det(h_1 h_2) = \det(h_1) \det(h_2) \quad,\quad
\det(\oline{h}) = \det(h) ,
\end{equation}
follow particularly easily. It is also known that $h$ has an inverse if and only if $\det(h)\neq 0$, in which case we can use the first of Formulas~(\ref{GAadjointIdent}), where we set $\uline{f}=h$, $A_r=h^{-1}(A)$ and $B_s=I$, to obtain an explicit expression
\begin{equation}
h^{-1}(A) = \frac{1}{\det(h)} \oline{h}(A I) I^{-1} ,
\end{equation}
where $A$ is a generic multivector.

The gauge field is an example of a linear function $a \mapsto h(a;q)$ that varies from point to point in the configuration space. Often, it is needed to form the derivative
\begin{equation}
b \cdot \dot{\partial}_q \dot{h}(a)
= b \cdot \dot{\partial}_q \dot{h}(\dot{a}) - b \cdot \dot{\partial}_q h(\dot{a}) ,
\end{equation}
which differentiates only the $q$-dependency of $h$, and not an eventual change of the vector field $a(q)$. 

When shall omit the scalar object ``$b \cdot \dot{\partial}_q$" when writing differential identities, if it is clear that the remaining dots are understood as directional derivatives in a generic direction. For example, the latter equation then acquires the neat form $\dot{h}(a) = \dot{h}(\dot{a}) - h(\dot{a})$.

Since $h h^{-1}$ is the identity for each point $q$,
\begin{equation} \label{GAhhDer}
\dot{h} h^{-1} = - h \dot{h^{-1}} .
\end{equation}
Moreover, the outermorphism $h$ satisfies a Leibniz rule
\begin{equation}
\dot{h}(A \wedge B) = \dot{h}(A) \wedge \dot{h}(B)
= \dot{h}(A) \wedge h(B) + h(A) + \dot{h}(B) ,
\end{equation}
which can be used to prove the property~(\ref{FieldStrMulti}) of the gauge field strength $F$:
\begin{align}
F(A_r \wedge B_s) 
&= -\oline{h} \big( \dot{\partial}_q \wedge \dot{\oline{h^{-1}}} (A_r \wedge B_s) \big) \nonumber\\
&= -\oline{h} \big( \dot{\partial}_q \wedge \dot{\oline{h^{-1}}} (A_r) \big) \wedge B_s 
- \oline{h} \big( \dot{\partial}_q \wedge \oline{h^{-1}} (A_r) \wedge \dot{\oline{h^{-1}}}(B_s) \big) \nonumber\\
&= F(A_r) \wedge B_s + (-1)^r A_r \wedge F(B_s) .
\end{align}
This can be iterated to yield
\begin{align} \label{FieldStrMulti2}
F(\blade{b}{r}) 
&= F(b_1) \wedge b_2 \wedge \ldots \wedge b_r
- b_1 \wedge F(b_2 \wedge \ldots \wedge b_r) \nonumber\\
&= \sum_{j=1}^r (-1)^{j-1} F(b_j) \wedge \bladecheck{b}{r}{j} .
\end{align}
(The ``checked" vectors are missing from the expression.)

Finally, let us recall the transformation rules (\ref{TrRules}), (\ref{TrGaugeFieldAll}) and (\ref{GCTrDeriv}) for $P$, $h$ and $\partial_q$, respectively, and check that $F(\bar{P})$ (with $\bar{P} \equiv \oline{h}(P)$) is invariant under gauge transformations. We have 
\begin{equation} \label{FieldStrTrPf}
F'(\bar{P}') 
= -\oline{h}\,\oline{f} \big( \oline{f^{-1}}(\dot{\partial}_{q}) \wedge \dot{\oline{f^{-1}}} \, \dot{\oline{h^{-1}}} (\bar{P}) \big)
= F(\bar{P}) - \oline{h}\big( \dot{\partial}_{q} \wedge \oline{f} \, \dot{\oline{f^{-1}}} (P) \big) .
\end{equation}
But the second term on the right-hand side vanishes on account of Eq.~(\ref{GCadjointCurl}), since
\begin{equation}
\dot{\partial}_{q} \wedge \oline{f} \, \dot{\oline{f^{-1}}}(P)
= -\dot{\partial}_{q} \wedge \dot{\oline{f}} \, \oline{f^{-1}}(P)
= 0 . 
\end{equation}

\subsubsection{Bivectors and antisymmetric transformations}
\label{sec:GAGCBivectAntisym}

A linear transformation $h$ is said to be \emph{antisymmetric} if it satisfies any of the equivalent conditions
\begin{equation} \label{AntisymConds}
\oline{h} = -h \quad\Leftrightarrow\quad
b \cdot h(a) = - a \cdot h(b) \quad (\forall a, b \in V) \quad\Leftrightarrow\quad
a \cdot h(a) = 0 \quad (\forall a \in V) .
\end{equation}
(Note that the first two equations can be obtained from the third one promptly by differentiation with respect to $a$.)
Any bivector $A$ gives rise to an antisymmetric mapping
\begin{equation}
\mathsf{A}(a) = a \cdot A = -A \cdot a ,
\end{equation}
and conversely, any antisymmetric mapping $h$ can be characterized by a bivector, namely, ${\frac{1}{2} e^j \wedge h(e_j)}$ (summation over $j$), since
\begin{equation}
a \cdot \big( e^j \wedge h(e_j) \big) 
= a \cdot e^j h(e_j) - e^j a \cdot h(e_j)
= 2 h(a) .
\end{equation}

Commutator of two antisymmetric mappings is again antisymmetric, and its characteristic bivector is generated by the commutator product (\ref{GAGCComProd}),
\begin{equation} \label{CommutRel}
[\mathsf{A}_1,\mathsf{A}_2](a)
= (\mathsf{A}_1 \mathsf{A}_2 - \mathsf{A}_2 \mathsf{A}_1)(a)
= (a \cdot A_2) \cdot A_1 - (a \cdot A_1) \cdot A_2
= a \cdot (A_2 \times A_1) .
\end{equation}
The algebra of antisymmetric matrices with the product $[.,.]$ can therefore be represented as the algebra of corresponding bivectors with the commutator product ``$\times$".

\subsubsection{Rotations and rotors}
\label{sec:GAGCRot}

In geometric algebra, a rotation $\mathsf{R}$ is conveniently represented using a \emph{rotor} $R$ as (see Refs.~\cite{ZatlSymConsLaws}, \cite{Hestenes} or \cite{DoranLas})
\begin{equation}
\mathsf{R}(a) = R a \rev{R} \quad,\quad R = \pm e^{-B/2} ,
\end{equation}
where $B$ is a bivector. This exponential form of rotors is ensured only in spaces with Euclidean or Lorentz signature (see \cite[Ch.~3-8]{Hestenes} or \cite{Riesz}), which we shall thus assume in this subsection. (The $\pm$-sign reflects the double-covering nature of the rotor representation, and has no effect on the expression for rotation, $R a \rev{R}$.)

Since bivectors change sign under reversion, $\rev{B} = -B$, and henceforth $\rev{R} = \pm e^{B/2}$, rotors enjoy the property
\begin{equation} \label{RotorId}
R \rev{R} = \rev{R} R = 1 .
\end{equation}
For completeness we indicate also inverse and adjoint rotations:
\begin{align}
\mathsf{R}^{-1}(a) = \rev{R} a R , \nonumber\\
\oline{\mathsf{R}}(b) = \rev{R} b R , \nonumber\\
\oline{\mathsf{R}^{-1}}(b) = R b \rev{R} .
\end{align}

When rotor-valued functions $R(q)$ on the configuration space $\mathcal{C}$ are considered, differentiation of relation (\ref{RotorId}) yields
\begin{equation} \label{RotorDif}
\dot{R} \rev{R} 
= - R \dot{\rev{R}} ,
\end{equation}
which is, importantly, a bivector. This fact allows to represent the Lie algebra of antisymmetric matrices $\mathfrak{so}(n)$, corresponding to the Lie group $SO(n)$ of rotations $\mathsf{R}$, as the algebra of bivectors (see Eq.~(\ref{CommutRel}), and for more details Ref.~\cite[Ch.~11.3]{DoranLas}).

Relation~(\ref{RotorDif}) can be used to obtain
\begin{align} \label{RotorDif2}
\dot{R} b \dot{\rev{R}}
= \dot{R} \rev{R} (R b \rev{R}) - (R b \rev{R}) \dot{R} \rev{R} 
= (2 \dot{R} \rev{R}) \cdot (R b \rev{R})
= (R b \rev{R}) \cdot (2 R \dot{\rev{R}}) ,
\end{align}
a formula employed in our study of the Yang-Mills gauge field in Sec.~\ref{sec:ExYM}.

\subsubsection{Shear mappings}
\label{sec:GAGCShear}

Elementary \emph{shear} transformation is a linear mapping defined by two orthogonal vectors $u$ and $v$,
\begin{equation}
\mathsf{S}_{u v}(a) := a + u \, v \cdot a \quad,\quad u \cdot v = 0 .
\end{equation}
During our treatment of the Yang-Mills gauge field in Sec.~\ref{sec:ExYM} we note that $h_{\rm YM^*}$ has slightly more general structure
\begin{equation} \label{GenShear}
\mathsf{S}(a;\veclist{u}{r};\veclist{v}{r}) = a + \sum_{j=1}^r u_j \, v_j \cdot a 
\quad,\quad u_j \cdot v_k = 0 \quad \forall j,k=1,\ldots,r ,
\end{equation}
which can, nevertheless, be written as a composition of elementary shears,
\begin{equation}
\mathsf{S}(a) = \mathsf{S}_{u_r v_r}\big( \ldots \mathsf{S}_{u_1 v_1}(a)\big) .
\end{equation}

Due to orthogonality of vectors $u_j$ and $v_k$, it is easy to form the inverse and the adjoint of the generic shear transformation (\ref{GenShear}),
\begin{align}
\mathsf{S}^{-1}(a) &=  a - \sum_{j=1}^r u_j \, v_j \cdot a  , \nonumber\\
\oline{\mathsf{S}}(b) &= b + \sum_{j=1}^r b \cdot u_j \, v_j   , \nonumber\\
\oline{\mathsf{S}^{-1}}(b) &=  b - \sum_{j=1}^r b \cdot u_j \, v_j .
\end{align}

\section{The field strength as torsion} \label{sec:FieldStrTorsion}

Let $\{ \gamma_j \}_{j=1}^{D+N}$ denote a basis (or \emph{frame}) in the tangent space of $\mathcal{C}$, and $\{ \gamma^j \}_{j=1}^{D+N}$ its reciprocal. Let us assume that this frame is constant, i.e., $q$-independent, and introduce an additional, possibly $q$-dependent, \emph{coordinate frame} $\{ e_\mu \}_{\mu=1}^{D+N}$, for which the Lie bracket between any pair of vectors vanishes,
\begin{equation} \label{CoordFrameLieBr}
[ e_\mu, e_\nu ] = e_\mu \cdot \partial_q e_\nu - e_\nu \cdot \partial_q e_\mu = 0 .
\end{equation}
Components of the gauge field with respect to these frames form the vielbein and its inverse (c.f. Appendix~C in Ref.~\cite{Lasenby1998})
\begin{equation}
h_j^{~\mu} = \gamma_j \cdot \oline{h}(e^\mu) \quad,\quad
h^j_{~\mu} = \gamma^j \cdot h^{-1}(e_\mu) .
\end{equation}

For the sake of correspondence, we shall identify the space of partial observables $\mathcal{C}$ with the spacetime of the \emph{teleparallel} theory of gravity \cite{Andrade2000,Aldrovandi}, where the vielbeins are used to define the \emph{Weitzenb\"{o}ck} (or \emph{affine}) connection with coefficients $h_j^{~\rho} \partial_\nu h^j_{~\mu}$. The gravitational field strength of teleparallel gravity is determined by the \emph{torsion} tensor, which is the antisymmetric part of the connection. Its components are calculated as follows:
\begin{equation}
T^\rho_{~\mu\nu}
= h_j^{~\rho} \partial_\mu h^j_{~\nu} - h_j^{~\rho} \partial_\nu h^j_{~\mu} 
= e_\mu \cdot \dot{\partial}_q \, \oline{h}(e^\rho) \cdot \dot{h^{-1}}(\dot{e}_\nu) 
- e_\nu \cdot \dot{\partial}_q \, \oline{h}(e^\rho) \cdot \dot{h^{-1}}(\dot{e}_\mu) , 
\end{equation}
and hence, in view of Eq.~(\ref{CoordFrameLieBr}),
\begin{equation}
T^\rho_{~\mu\nu}
= (e_\nu \wedge e_\mu) \cdot \big( \dot{\partial}_q \wedge \dot{\oline{h^{-1}}} \, \oline{h}(e^\rho) \big)
= h^{-1}(e_\mu \wedge e_\nu) \cdot F(\oline{h}(e^\rho)) ,
\end{equation}
where $F$ is the gauge field strength defined in Eq.~(\ref{FieldStrength}).


\end{document}